\documentclass[12pt]{article}
\usepackage{fullpage}
\usepackage{amsmath}
\usepackage{amssymb}
\usepackage[dvips]{epsfig}

\def\##1{\underline{#1}}
\def\=#1{\underline{\underline{#1}}}

\def\+#1{\underline{\bf #1}}
\def\*#1{\underline{\underline{\bf #1}}}

\def\le{\left(}
\def\ri{\right)}
\def\les{\left[}
\def\ris{\right]}

\def\.{\mbox{ \tiny{$^\bullet$} }}

\begin{document}

\begin{center}

{\bf {\Large A short note on climbing gravity improbable :\\ Superradiant ring fellowship of launching\\ the high Lorentz-factor outflows/jets}}

 \vspace{4mm} \large

Sandi Setiawan\footnote{Fax: + 44 131 650 6553; e--mail:
S.Setiawan@ed.ac.uk}\\
{\em School of Mathematics,
University of Edinburgh, Edinburgh EH9 3JZ, UK}\\

\end{center}

\vspace{4mm}

\normalsize

\begin{abstract}
\noindent 
The spin energy extraction from a rotating black hole by amplification 
of scalar waves inside superradiance resonance wedge cavity (ring), 
followed by jet formation via magnetic reconnection assisted by 
the amplified scalar fields, is proposed. 
This mechanism may explain the availability of energy for as well as 
the formation of relativistic jets with high Lorentz-factor 
in microquasars and quasars. 
It is speculated that it may also explain 
that the spin of the central black hole, 
which facilitates the formation of superradiance cavity, 
which in turn would produce 
the amplified scalar waves by taking the hole's spin energy, 
is the source for the distinction between 
radio-loud and radio-quiet active galaxies. 

\end{abstract}


\vspace{4mm}

\section{Introduction}

Recently it was suggested that superradiant scattering of magnetosonic waves in a toroidal magnetosphere around a Kerr black hole may play a significant role in providing the energy for gamma-ray bursts \cite{putten1}. Using the WKB approximation, the amplification of scalar waves confined in a equatorial cavity wedge was computed and it was found that the cavity which acts as a waveguide for magnetosonic waves may give rise to instabilities in all superradiant modes. The author then suggested that the result for scalar waves should generalize to electromagnetic waves. However, this was not the case, though it was found that the scalar waves were amplified more strongly \cite{aguirre1} than the results obtained in \cite{putten1}.

It is also well-known that the black hole rotational energy may be extracted via Penrose process \cite{penrose1,bardeen1,wilkins1,wald1} (see also \cite{christodoulou1,christodoulou2,teukolsky0,teukolsky1,teukolsky2,teukolsky3,chandra0,aliev1}). Recent simulations showed that the so-called magnetohydrodynamical (MHD) Penrose process of extracting spin energy of a rotating black hole is only true for a short period of time when a large fraction of magnetic fields anchored in the black hole ergosphere \cite{komissarov1}, contrary to the claim suggested previously that this process may play a quite important role \cite{koide1}. The magnetic field lines are quickly pulled out from the ergosphere and moved into the black hole, and the MHD Penrose process almost completely stopped to operate \cite{komissarov1}. However, the extraction of rotational energy of the black hole is still taking place via Blandford-Znajek mechanism \cite{blandford1,znajek1} (see also \cite{blandford2,znajek2,znajek3}; and \cite{wilson1,xinli1,wald2,king1,petterson1,petterson2,chitre1} as well as \cite{bhmag1,bhmag1a,bhmag2,bhmag3,bhmag4}), which is believed to be responsible for the creation of jets (see, for example, \cite{narayan1,semenov1} and related references therein), that may be originated from within a few Schwarschild radius from the central black hole \cite{wilms1,iwasawa1,dabrowski1}. Unfortunately, from the results of advanced recent simulations (see, for example, \cite{komissarov1}) we learn that relativistic as well as collimated jets are failed to attain, and one should look for other explanations. However, there are still claims and counter-claims going on about the subject (see, for example, \cite{komissarov1,koide1,shibata1,villiers1,villiers2,villiers3,mckinney1,macfadyen1,macfadyen2}).

It was also discovered recently that a rapidly rotating black hole may have the so-called `superradiance resonance cavity' just outside it \cite{andersson1,andersson2} (see also \cite{superrad1,superrad2}). The existence of this cavity may physically explain the presence of large number of slowly damped quasi-normal modes (QNMs) trapped inside it \cite{andersson1}. If this kind of cavity happens to be physically similar to the wedge cavity in \cite{putten1}, in which it is proposed that it resembles the vacuum cavity around a black hole enclosed by a mirror as found in \cite{press1}, then it would be worth exploring the possibility of extracting spin energy from a rotating black hole via amplification of scalar waves trapped inside the cavity, which then followed by jet formation via magnetic reconnection assisted by those amplified scalar fields.

\section{Theoretical framework}

Massive scalar fields in Kerr spacetime background are unstable and their amplitudes can grow without bound \cite{detweiler1,strafuss1}. However, the instability growth time is too short to be astrophysically important \cite{detweiler1}. For the massless scalar fields, which are very important in the proposed scenario explored in this paper, one can find interesting discussion in \cite{press1}; scalar fields as dark energy which produce accelerating environment can be found in \cite{scalar1,scalar2,scalar3} (see \cite{scalar4,scalar5} for scalar fields in the context of inflationary universe model; see also \cite{weinberg0} for general account on scalar fields and \cite{simu} for a version of its simulations). Related to this topic, one can also find earlier accounts on this subject in \cite{starobinskii1,starobinskii2,zeldovich1}. 

We will not do detailed analysis in this paper. What will be done here is only to propose a theoretical framework on which we may build up a scenario about what physical mechanism that may be responsible for launching the high Lorentz-factor outflows/jets. First, we consider a rotating Kerr black hole. The Kerr black hole would transform itself into the Kerr-Newman hole if it is immersed in the magnetic fields since the induced charges would be acquired in this situation (see, for example, \cite{xinli1,wald2}). We then investigate the perturbation of massless scalar fields in this Kerr-Newman spacetime \cite{teukolsky0,teukolsky1,teukolsky2,teukolsky3} (see also \cite{fackerell,seidel}). According to the authors of \cite{andersson1,andersson2}, superradiance resonance cavity would be present in this situation and it is located just outside the black hole's outer event horizon (see also \cite{superrad1,superrad2}). We compare the result obtained from that of Kerr-Newman spacetime with the one from a very narrow equatorial wedge background \cite{putten1,aguirre1}. We would then consider a kind of magnetized Kerr-Newman-de Siter spacetime (see, for example, \cite{stuchlik1,stuchlik2,stuchlik3,gh1,gibbons1,vasudevan1,suzuki1,suzuki2}) as the relevant background due to the presence of the amplified massless scalar waves caused by the superradiance cavity to investigate the motion of the particles/waves/radiation in the vicinity of the black hole.

The Kerr-Newman-de Sitter metric in the Boyer-Lindquist coordinates \cite{gh1,suzuki1,suzuki2} can be expressed as follows : 

\begin{eqnarray}
\label{bl}
ds^2 &=& -\rho^2
\left(\frac{dr^2}{\Delta_r}+\frac{d\theta^2}{\Delta_\theta}\right)
-\frac{\Delta_\theta \sin^2\theta}{(1+\frac{\displaystyle \Lambda a^2}{\displaystyle 3})^2 \rho^2}
[a \, dt-(r^2+a^2) \, d\varphi]^2 \nonumber \\ 
&&  +\frac{\Delta_r}{(1+\frac{\displaystyle \Lambda a^2}{\displaystyle 3})^2 \rho^2}(dt-a\sin^2\theta \, d\varphi)^2,
\end{eqnarray}
with 
\begin{eqnarray}
\Delta_r &=& \le r^2+a^2 \ri \le 1-\frac{\Lambda r^2}{3} \ri -2Mr+ Q^2 \,, \\
\Delta_\theta &=& 1+\frac{\Lambda}{3} a^2 \cos^2\theta \,, \\
\rho^2 &=& r^2 + a^2 \cos^2\theta \,, 
\end{eqnarray}
wherein $\Lambda$ is the term (that behaves like `cosmological constant') which represents the amplified uncharged massless scalar fields, $M$ the mass of the black hole, $a$ its angular momentum per unit mass and $Q$ its charge. However, it is very unlikely that $\Lambda$ would be constant in this case since the scalar fields may be produced differently depending on their distance from the hole's outer event horizon (perhaps, for example, $\Lambda \propto \frac{\displaystyle 1}{\displaystyle r}$), and the scalar fields themselves would also evolve with time.

Since we are dealing with a Kerr rotating black hole immersed in magnetic fields, we would consider the Kerr-Newman background because the Kerr black hole would acquire the induced charge caused by the magnetic fields. We then investigate separated equations for the perturbation of uncharged massless fields with arbitrary spin $s$ in Kerr-Newman background using Boyer-Lindquist coordinates. We can simply take $\Lambda = 0$ in this case. And, we get the angular Teukolsky equation \cite{teukolsky0} (see also \cite{aguirre1}) :

\begin{equation}
\label{eqn:sx}
\les \frac{1}{\sin\theta}\frac{d}{d\theta}\sin\theta \frac{d}{d\theta} + a^2\omega^2\cos^2\theta + 2 a \omega s \cos\theta -\frac{(m+s\cos\theta)^2}{\sin^2\theta} + E-s^2  \ris S = 0 \,,
\end{equation}
and the radial Teukolsky equation is given by \cite{teukolsky0} (see also \cite{aguirre1}) :
\begin{equation}
\les \Delta^{-s}\frac{d}{dr}\Delta^{s+1}\frac{d}{dr} +
  \frac{K^2+2is(r-M)K}{\Delta} -4ir\omega s 
-E-2am\omega-a^2\omega^2+ s(s+1) \ris R = 0 \,, 
\end{equation}
wherein $\Delta \equiv r^2 - 2Mr + a^2 + Q^2$ and $K \equiv \le r^2+a^2 \ri \omega + am$, with $E$ an angular eigenvalue.

Chandrasekhar showed that the radial equation can be expressed as a wave equation with a potential barrier \cite{chandra0}. In this paper, we are interested in the uncharged massless scalar fields, i.e., $s=0$. For this case, the potential barrier is given by :

\begin{equation}
V = \frac{\Delta}{\rho^4} \les \lambda + \frac{1}{\rho^2} \le \Delta + 2r \le r-M \ri \ri -3 \frac{r^2\Delta}{\rho^4} \ris \,,
\end{equation}
and $\lambda$ is the angular eigenvalue given by $\lambda = E+a^2\omega^2+2am\omega$, in which $E$ is \cite{fackerell,seidel} (see also \cite{aguirre1}) :

\begin{equation}
E = l(l+1) + 2a^2\omega^2 \les \frac{m^2-l(l+1)+\frac{1}{2}}{(2l-1)(2l+3)} \ris + O\les (a\omega)^4 \ris \,,
\end{equation} 
which is obtained from the boundary conditions of regularity at $\theta=0$ and $\theta=\pi$ (see \cite{aguirre1}). This value should then be compared with the one obtained by assuming a constant angular function across the narrow superradiance resonance wedge cavity, which is 
\begin{equation}
E \simeq m^2 + s^2 \,,
\end{equation}
by taking $dS/d\theta=0$ in equation (\ref{eqn:sx}), followed by taking the approximation $\cos\theta \rightarrow 0$ and $\sin\theta \rightarrow 1$ (see \cite{putten1} and also \cite{aguirre1}). 

For the scalar case where $s=0$, 
the very narrow wedge geometry changes the angular eigenvalue for $m=1$
from $E = 2+O[(a\omega)^2]$ to $E=1$; the eigenvalue would add to the
height of the potential barrier, so the lower eigenvalue in the wedge
geometry would produce much stronger superradiance \cite{aguirre1}. 
We should also note that the presence of the charge $Q$ reduces the 
size of the outer event horizon and the static limit of the hole.

We should then be able to analyze the motion of the particles/waves/radiation 
in the vicinity of the black hole immersed in magnetic fields (which may be 
driven by the currents in the accretion disk surrounded the hole) 
using a more general magnetized Kerr-Newman-de Sitter spacetime 
(see, for example, \cite{bhmag1a}; see also \cite{kraniotis1}) with 
$\Lambda$ inconstant, in order to establish the right environment for 
launching a relativistic outflows/jets. Detailed accounts will be 
published elsewhere \cite{sandi2}.

We note that 
this superradiance effect would take the spin energy of the black hole and 
create the background of (amplified) scalar fields which is similar to 
the de Sitter spacetime (these scalar fields may also be interpreted as
something similar to dark energy in an accelerating de Sitter-like universe
-- see, for example, \cite{scalar1,scalar2,scalar3}). 
Combining this background with the spacetime created by 
the charged-induced rotating black hole \cite{xinli1} 
would result in a spacetime with a kind of 
Kerr-Newman-de Sitter metric especially in a region close to the black hole. 
The difficulties of having relativistic flows and jets are solved 
by the fact that it would be natural for matter and radiation near 
the black hole to attain high-Lorentz factor velocity by `riding' 
the amplified massless scalar waves created 
in the ergosphere of a rotating black hole
which provide the de Sitter-like spacetime environment which is 
accelerating. 
This process may explain the availability of energy for as well as 
the formation of relativistic jets with very high-Lorentz factor 
in microquasars 
(for example, in form of gamma-ray bursts \cite{sandi}) and quasars \cite{rees}. 

In addition, the generated magnetic reconnections (in relation with 
the matter of the accreted disk) may also help to collimate the jets. 
This mechanism may also explain that the spin of the central black hole, 
which facilitates the formation of superradiance cavity, 
which in turn would produce the amplified uncharged massless 
scalar fields by taking the hole's spin energy, 
is the source for the distinction between 
radio-loud and radio-quiet active galaxies \cite{wilson1}. 
It may also be responsible for the presence of the dark energy 
(or/and, perhaps, dark matter) in our universe.
We also note that the scenario proposed here is different from that of 
\cite{reva1}.

\section{Concluding remarks}

The proposal considered in this paper is still in its very early development. Further consideration, both analytic and numerical, needs to be explored in order to confirm the claims made in this paper. It is very tempting to know what large-scale numerical simulations would say about this by considering the full magnetohydrodynamical general relativistic approach and by also taking into account the effects arise from the presence and evolution of superradiance wedge cavity (ring) in the ergosphere of a rotating black hole.

\bigskip

\noindent{\bf Acknowledgement}

\noindent The author acknowledges EPSRC for support under grant GR/S60631/01.


\end{document}